\documentclass[journal=ancac3,manuscript=article]{achemso}
\usepackage{graphics} 
\usepackage{graphicx} 
\usepackage{amsmath}
\usepackage{amssymb}
\usepackage{soul}
\usepackage{pdfpages}

\author{Omar Khatib}
\email{Omar.Khatib@Colorado.edu}
\affiliation{Department of Physics, University of California, San Diego, La Jolla, California 92093, USA}
\alsoaffiliation{Department of Physics, Department of Chemistry, and JILA, University of Colorado, Boulder, Colorado 80309, USA}
\author{Joshua D. Wood}
\affiliation{Department of Materials Science and Engineering, Northwestern University, Evanston, Illinois 60208, USA} 
\alsoaffiliation{Department of Electrical and Computer Engineering, University of Illinois at Urbana-Champaign, Urbana, Illinois 61801, USA}
\alsoaffiliation{Beckman Institute for Advanced Science and Technology, University of Illinois at Urbana-Champaign, Urbana, Illinois 61801, USA}
\alsoaffiliation{Micro and Nanotechnology Laboratory, University of Illinois at Urbana-Champaign, Urbana, Illinois 61801, USA}
\author{Alexander S. McLeod}
\affiliation{Department of Physics, University of California, San Diego, La Jolla, California 92093, USA}
\author{Michael D. Goldflam}
\affiliation{Department of Physics, University of California, San Diego, La Jolla, California 92093, USA}
\author{Martin Wagner}
\affiliation{Department of Physics, University of California, San Diego, La Jolla, California 92093, USA}
\author{Gregory L. Damhorst}
\affiliation{Micro and Nanotechnology Laboratory, University of Illinois at Urbana-Champaign, Urbana, Illinois 61801, USA}
\alsoaffiliation{Department of Bioengineering, University of Illinois at Urbana-Champaign, Urbana, Illinois 61801, USA}
\author{Justin C. Koepke}
\affiliation{Department of Electrical and Computer Engineering, University of Illinois at Urbana-Champaign, Urbana, Illinois 61801, USA}
\alsoaffiliation{Beckman Institute for Advanced Science and Technology, University of Illinois at Urbana-Champaign, Urbana, Illinois 61801, USA}
\alsoaffiliation{Micro and Nanotechnology Laboratory, University of Illinois at Urbana-Champaign, Urbana, Illinois 61801, USA}
\author{Gregory P. Doidge}
\affiliation{Department of Electrical and Computer Engineering, University of Illinois at Urbana-Champaign, Urbana, Illinois 61801, USA}
\alsoaffiliation{Beckman Institute for Advanced Science and Technology, University of Illinois at Urbana-Champaign, Urbana, Illinois 61801, USA}
\alsoaffiliation{Micro and Nanotechnology Laboratory, University of Illinois at Urbana-Champaign, Urbana, Illinois 61801, USA}
\author{Aniruddh Rangarajan}
\affiliation{Department of Electrical and Computer Engineering, University of Illinois at Urbana-Champaign, Urbana, Illinois 61801, USA}
\alsoaffiliation{Beckman Institute for Advanced Science and Technology, University of Illinois at Urbana-Champaign, Urbana, Illinois 61801, USA}
\alsoaffiliation{Micro and Nanotechnology Laboratory, University of Illinois at Urbana-Champaign, Urbana, Illinois 61801, USA}
\author{Rashid Bashir}
\affiliation{Department of Electrical and Computer Engineering, University of Illinois at Urbana-Champaign, Urbana, Illinois 61801, USA}
\alsoaffiliation{Micro and Nanotechnology Laboratory, University of Illinois at Urbana-Champaign, Urbana, Illinois 61801, USA}
\alsoaffiliation{Department of Bioengineering, University of Illinois at Urbana-Champaign, Urbana, Illinois 61801, USA}
\author{Eric Pop}
\affiliation{Department of Electrical Engineering, Stanford University, Stanford, California 94305, USA}
\author{Joseph W. Lyding}
\affiliation{Department of Electrical and Computer Engineering, University of Illinois at Urbana-Champaign, Urbana, Illinois 61801, USA}
\alsoaffiliation{Beckman Institute for Advanced Science and Technology, University of Illinois at Urbana-Champaign, Urbana, Illinois 61801, USA}
\alsoaffiliation{Micro and Nanotechnology Laboratory, University of Illinois at Urbana-Champaign, Urbana, Illinois 61801, USA}
\author{Mark H. Thiemens}
\affiliation{Department of Chemistry and Biochemistry, University of California, San Diego, La Jolla, California 92093, USA }  
\author{Fritz Keilmann}
\affiliation{Ludwig-Maximilians-Universit\"at and Center for Nanoscience, 80539 M\"unchen, Germany}  
\author{D. N. Basov}
\affiliation{Department of Physics, University of California, San Diego, La Jolla, California 92093, USA}  

\title{Graphene-Based Platform for Infrared Near-Field Nanospectroscopy of Water and Biological Materials in an Aqueous Environment}

\begin{document}

\begin{abstract}
Scattering scanning near-field optical microscopy (\textit{s}-SNOM) has emerged as a powerful nanoscale spectroscopic tool capable of characterizing individual biomacromolecules and molecular materials. However, applications of scattering-based near-field techniques in the infrared (IR) to native biosystems still await a solution of how to implement the required aqueous environment. In this work, we demonstrate an IR-compatible liquid cell architecture that enables near-field imaging and nanospectroscopy by taking advantage of the unique properties of graphene. Large-area graphene acts as an impermeable monolayer barrier that allows for nano-IR inspection of underlying molecular materials in liquid. Here, we use \textit{s}-SNOM to investigate the tobacco mosaic virus (TMV) in water underneath graphene. We resolve individual virus particles and register the amide I and II bands of TMV at ca. 1520 and 1660 cm$^{-1}$, respectively, using nanoscale Fourier transform infrared spectroscopy (nano-FTIR). We verify the presence of water in the graphene liquid cell by identifying a spectral feature associated with water absorption at 1610 cm$^{-1}$. 
\end{abstract}

In biological and life sciences, Fourier transform infrared (FTIR) spectroscopy serves as a ubiquitous noninvasive probe of vibrational fingerprints used to identify chemical compounds and molecular species.\cite{Movasaghi2008} This information is the basis for nonperturbative and label-free analysis of cell functionality.\cite{Baker2014} For example, small changes in frequencies and line shapes of IR absorption bands due to specific proteins or protein conformations can be used to characterize cells and tissues linked to diseases such as Alzheimer's\cite{Juszczyk2009} and cancer.\cite{Movasaghi2008,Ami2013} Detailed databases document molecular absorption bands of biologically relevant systems,\cite{Movasaghi2008} whereas elaborate data processing schemes strive to minimize confounding effects associated with the exploration of realistic biological materials.\cite{Baker2014}  Despite the widespread use of FTIR spectroscopy and microscopy in biological sciences, these experimental methods suffer from a number of fundamental shortcomings. With a typical IR absorption coefficient for molecules of ca. 1000 cm$^{-1}$, aggregates of $\sim$ 1 $\mu$m depth are needed to assess the spectral features of biological systems.\cite{Gucciardi2009} Further, \textit{in vivo} studies of these systems require characterization in an aqueous environment. However, the strong absorption from the vibrational and rotational modes of liquid water presents a large undesirable background, alleviated in some cases by substitution of deuterium oxide and/or specially prepared buffers.\cite{Haris1994} Lastly, the long wavelength of IR light represents a limit to spatial resolution imposed by diffraction, and thus does not allow for characterization of any subcellular components individually.

Scattering scanning near-field optical microscopy (\textit{s}-SNOM) allows for the collection of IR spectra from pixels as small as 10 x 10 nm$^2$, irrespective of the infrared wavelength involved.\cite{Hillenbrand2002a} In an \textit{s}-SNOM apparatus, the diffraction limit is circumvented by combining the use of an atomic force microscope (AFM) with IR lasers enabling near-field imaging.\cite{Hillenbrand2001,Keilmann2004,Atkin2012} Incident light illuminates the sharp conducting tip in the vicinity of a sample. Scattered light from the coupled tip-sample system is detected in the far field and carries dielectric information from the local near-field interaction. The incident field is highly confined to the tip apex, resulting in an IR or optical probe with spatial resolution limited only by the tip diameter and not by the wavelength. Coupling monochromatic light to \textit{s}-SNOM allows acquisition of single-frequency near-field images, while use of a broadband coherent source\cite{Keilmann2012}  enables nanoscale Fourier-transform IR spectroscopy (nano-FTIR).\cite{Amarie2009,Huth2011,Dai2014}

\begin{figure*}
\centering
\includegraphics[hiresbb=true, width=8.6cm]{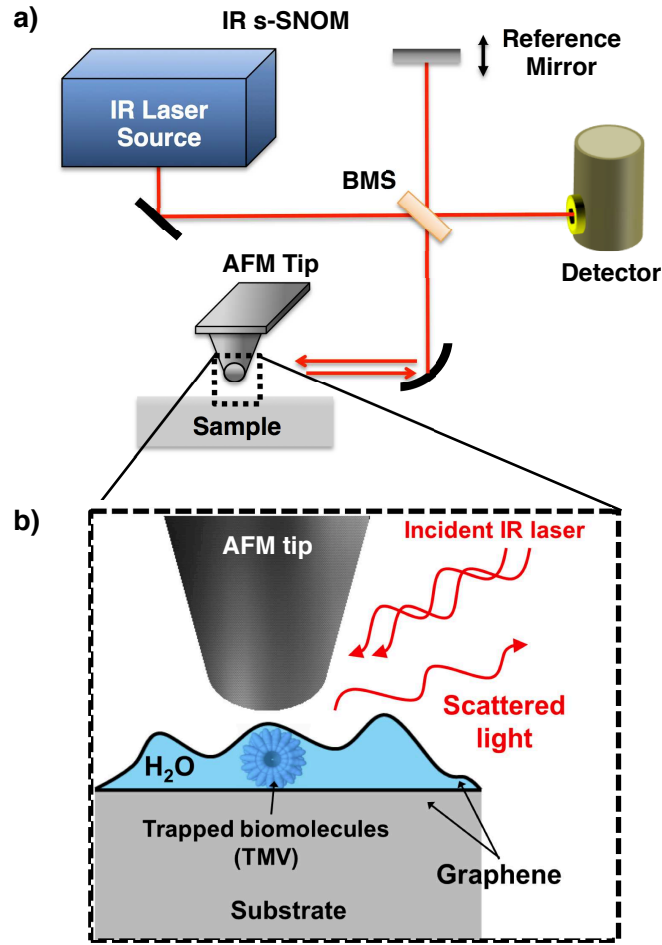}
\caption{"Wet" \textit{s}-SNOM setup with an IR-compatible graphene liquid cell. (a) Schematic of scattering-based near-field IR imaging and spectroscopy apparatus. A continuous wave CO$_2$ laser is used for monochromatic nanoimaging, and pseudoheterodyne detection with the reference mirror oscillating at 300 Hz. For nano-FTIR, a coherent broadband mid-IR continuum source is used in an asymmetric Michelson interferometer configuration, where one arm measures the backscattering from the AFM tip. (b) Geometry of the tip-sample interaction, sketching virus particles in water trapped beneath a large-area CVD graphene sheet.}
\label{schem}
\end{figure*}

Various interferometric detection schemes can be employed to extract both amplitude and phase of the near-field scattered signal.\cite{Schnell2014,Gucciardi2008} A schematic of the typical \textit{s}-SNOM experimental setup is shown in Figure \ref{schem}a. IR nanoimaging\cite{Brehm2006} and nanospectroscopy\cite{Amenabar2013} have demonstrated the ability to resolve individual subcellular biological components and chemically identify specific proteins.\cite{Berweger2013} However, measurement in liquid water is a precondition for studying living systems. As the IR absorption of water is strong, a general applicability of IR \textit{s}-SNOM and nano-FTIR spectroscopy in biology necessitates novel means to provide the required physiological conditions. Another complication of nano-IR spectroscopy in aqueous environments stems from the difficulties associated with the dynamics of tapping-mode AFM in fluids. This latter mode is required for quantitative nano-IR measurements that rely on rejection of far-field scattering and also of various topographic artifacts based on higher-harmonic demodulation protocols\cite{Keilmann2004}. However, harmonic motion of the cantilever immersed in water is difficult to implement.\cite{Legleiter2005,Preiner2007}  

Here we propose an IR-compatible liquid cell architecture for \textit{s}-SNOM and nano-FTIR, enabling near-field imaging and spectroscopy by taking advantage of the unique properties of graphene. Recent work has demonstrated the ability to trap liquids beneath graphene,\cite{Xu2010,He2012,Wood2012} an atomically thin sheet of carbon atoms with remarkable mechanical, electrical, and optical properties.\cite{Novoselov2012,Basov2014} Graphene-based liquid cell architectures have already enabled nanoscale studies of biomaterials using high-resolution probes such as scanning tunneling microscopy (STM)\cite{He2012} and transmission electron microscopy (TEM).\cite{Wang2014} The IR transparency of graphene allows for extending the use of such cells to IR \textit{s}-SNOM. In this work, we investigate the tobacco mosaic virus (TMV), a prototypical biological standard, trapped in water underneath graphene.\cite{Wood2013} The large-area graphene acts as an impermeable monolayer lid that allows for nano-IR interrogation of the underlying molecular materials in an aqueous environment (Figure \ref{schem}b). We employ graphene liquid cells supported by either mica or SiO$_2$ substrates (see Materials and Methods). We resolve individual viruses through graphene, and observe anticipated contrasts in near-field amplitude and phase images. Further, we register spectroscopic resonances specific to TMV encapsulated in the liquid cell. Our work paves the way for future studies using scattering-based near-field IR spectroscopy on biological systems in aqueous media.

There has been considerable development of aperture-based SNOM for the visible range, partly using liquid cells. However, these methods are based on fluorescence contrast and thus require chemical sample modification.\cite{Zhong2009,Hu2009,Dickenson2010,Lewis1999,Edidin2001,Lange2001,Hinterdorfer2011,Kapkiai2004,Zanten2010,Dunn1999,Herrmann2009} IR has the advantage of requiring no label for chemical recognition, but cannot give submicrometer resolution with an aperture-based SNOM.

\section{Results and Discussion}

\begin{figure*}
\centering
\includegraphics[hiresbb=true,width=17cm]{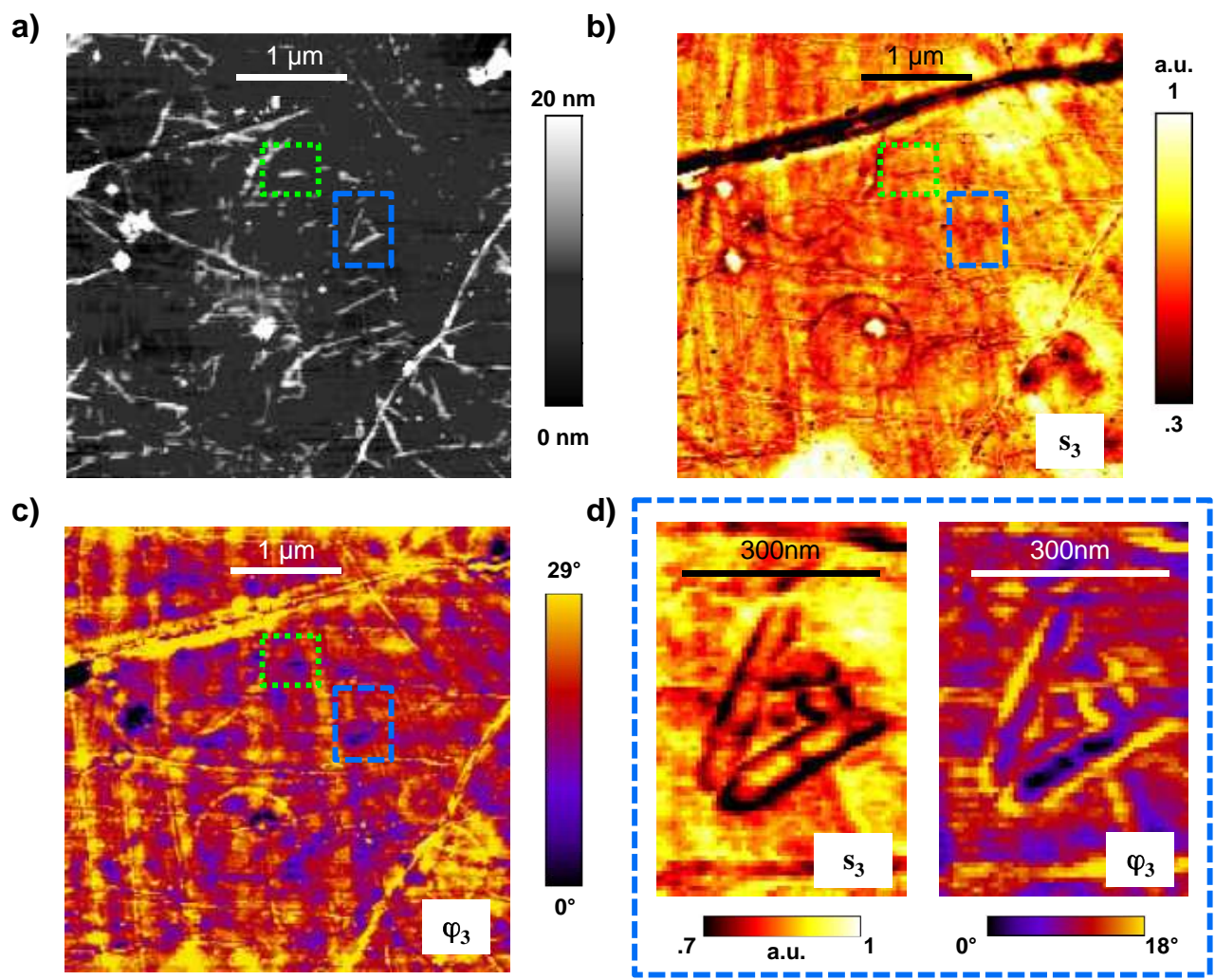}
\caption{IR \textit{s}-SNOM imaging at 890 cm$^{-1}$ of an aqueous solution on mica covered by an atomically thin graphene sheet: (a) AFM topography, (b) amplitude s$_3$, and (c) phase $\phi_3$ image of IR backscattering defined in the text and (d) high-resolution amplitude s$_3$ and phase $\phi_3$ images of the blue dashed boxed region in (a)-(c).} 
\label{imaging}
\end{figure*}

Tobacco mosaic viruses are rod shaped virions, nominally 300 nm in length and 18 nm in diameter. In dry preparations they are identifiable from AFM topography by their height and length, but usually exhibit an apparent width of ca. 50 nm.\cite{Brehm2006,Trinh2011,Alonso2013}  Figure \ref{imaging}a shows the topography of our \textit{s}-SNOM imaging of TMV contained with water under a graphene lid on mica. While long, thin features tentatively represent folds and wrinkles of graphene,\cite{Fei2013} smaller objects of different types abound. Some of them seem assignable to TMV from their shape and length, even though their height is only about 10 nm. Recent AFM studies have shown that strong attractive forces that bind graphene to the substrate indeed may deform TMV to below 10 nm height.\cite{Wood2013} Similar deformation and restructuring is to be expected for other residues from the processed tobacco plant and also for water.

Shown in Figure \ref{imaging}b and \ref{imaging}c are the \textit{s}-SNOM amplitude and phase images, respectively, acquired at 890 cm$^{-1}$ concurrently with the topography image in Figure \ref{imaging}a. Contrast in near-field images is indicative of the dielectric property of the material under the tip apex; dielectric resonances in the sample lead to significant phase $\phi_n$ and amplitude s$_n$ signatures in the backscattered infrared light, where n is a harmonic of the tip tapping frequency (see Materials and Methods).\cite{Hillenbrand2002,Taubner2004} Absorption is approximately equivalent to the product s$_n$ sin $\phi_n$, as we detail later.\cite{Huth2012} At the specific frequency used, we expect considerable contribution to the phase signal due to absorption from mica,\cite{Liang1998,Vedder1964} and somewhat weaker absorption due to water,\cite{Hale1973} PMMA,\cite{Huth2012} and protein.\cite{Movasaghi2008} Interestingly, the IR images (Figure \ref{imaging}b and \ref{imaging}c) show a network of high-phase objects with no counterpart in topography (Figure \ref{imaging}a). These include near circular structures with a somewhat diffuse boundary, as well as a vertical stripe pattern with approximately 100-150 nm width. These features, whose assignment requires a more extensive nano-FTIR investigation, are consistent with the presence of varying amounts of nanoconfined water, which has been shown to form filaments and conform to the templated CVD graphene morphology.\cite{Wood2013}

Many smaller objects show a lower IR phase than the background (which we tentatively assign to the mica substrate), with several objects revealing a topographical height of approximately 10 nm. Among these are graphene-encapsulated TMV (boxed regions in Figure \ref{imaging}a-c), with the width and length determined from the AFM topography, which we examine in more detail below.

\begin{figure*}
\centering
\includegraphics[hiresbb=true,width=8.6cm]{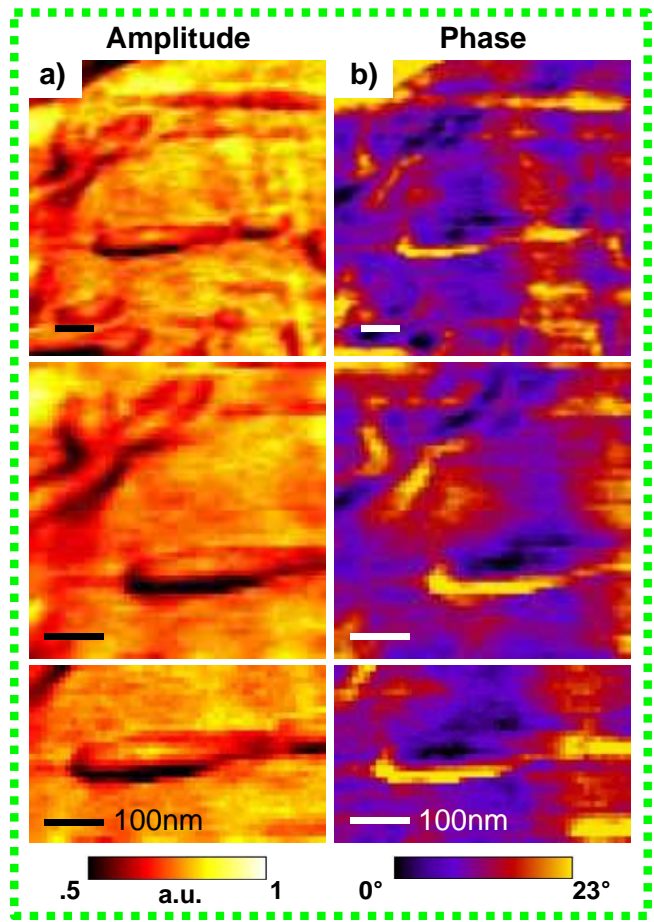}
\caption{Repeated IR \textit{s}-SNOM (a) amplitude s$_3$ and (b) phase $\phi_3$ images of green dotted boxed region shown in Fig \ref{imaging}a-c. All scale bars are 100 nm. } 
\label{reimaging}
\end{figure*}

High-resolution nano-IR images of apparently isolated TMV (colored boxes in Figure \ref{imaging}a) are shown in Figure \ref{imaging}d and Figure \ref{reimaging}, with a pixel size of 10 nm and the resolution limited by the tip radius. The near-field amplitude and phase at third harmonic demodulation are shown in the left and right panels of Figure \ref{imaging}d, respectively, representing the blue dashed boxed region of Figure \ref{imaging}a. We observe significant \textit{s}-SNOM contrast in regions with TMV. While optical contrast in the near-field scattering amplitude images can sometimes be highly correlated with the surface topography, phase shifts are less susceptible to such effects and in principle more sensitive to electronic resonances and vibrational absorptions.\cite{Huth2012} In addition, higher-harmonic signal components are progressively less sensitive to topographic artifacts, as a consequence of the highly nonlinear tip-sample near-field interaction.\cite{Keilmann2004} Interestingly, we also observe large contrast in the immediate regions bordering TMV in both amplitude and phase, indicating a variation in either the quantity or property of the water adjacent to the encapsulated viruses. The lower phase values over the TMV suggest that the dominant contribution to the near-field phase signal originates from the mica substrate or possibly the water layer. We further discuss the origin of the near-field contrast surrounding the TMV later.

Figure \ref{reimaging} shows several high-resolution \textit{s}-SNOM scans for the green dotted boxed region depicted in Figure \ref{imaging}a-c. The remarkable reproducibility of all salient features evident in both the near-field amplitude (Figure \ref{reimaging}a) and phase (Figure \ref{reimaging}b) images affirms IR \textit{s}-SNOM as a robust, nondestructive characterization tool for graphene-based liquid cell structures.

\begin{figure*}
\centering
\includegraphics[hiresbb=true,width=16cm]{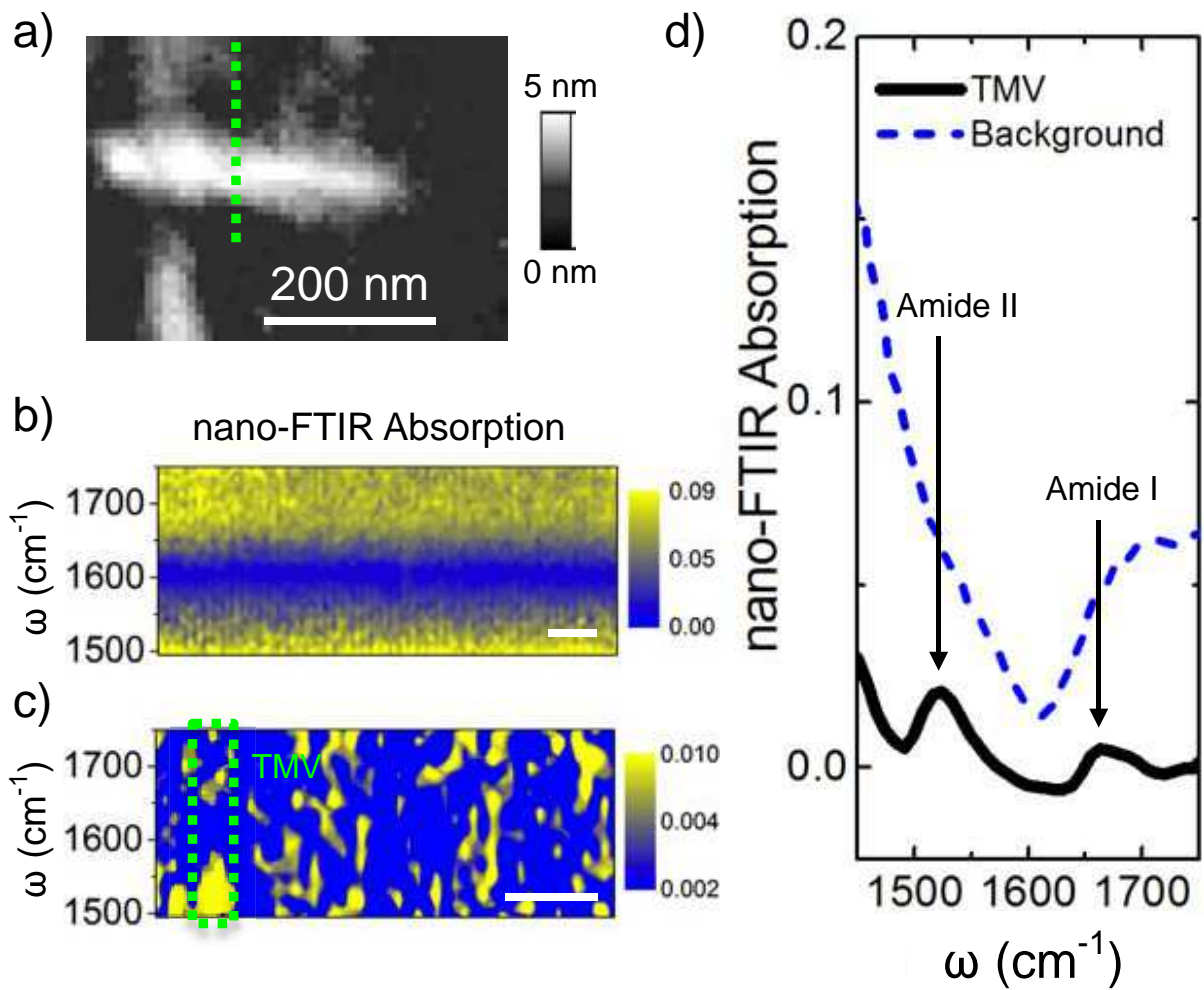}
\caption{Nano-FTIR absorption spectra of TMV in water taken with a graphene liquid cell on a SiO$_2$ substrate: (a) AFM topography showing two virions; (b) spectroscopic line scan in a featureless vicinity of virions showing 100 nano-FTIR absorption spectra acquired in 60 s each at 8 cm$^{-1}$ resolution, computed from both amplitude and phase signals at second order demodulation and normalized to Au; (c) spectroscopic line scan as in (b) but along the green dashed line crossing the virion in (a), showing 50 nano-FTIR absorption spectra after subtraction of the average background spectrum and plotted on a logarithmic color scale. All scale bars indicate a 200 nm length. (d) Average over all spectra in (b) shown as the blue dashed curve, with prominent spectral slopes assignable to water in the SiO$_2$-supported liquid cell, leaving a minimum near 1610 cm$^{-1}$; the virion spectrum averaged over the green boxed region in (c) is shown as the black curve.}
\label{ftir}
\end{figure*}

Using single-frequency IR nanoimaging, we have shown that in the near-field images we are able to identify individual virus particles surrounded by water through a monolayer graphene lid. Additionally, we collected nano-FTIR spectra of TMV and the local aqueous environment. For nanospectroscopy measurements, we found that mica substrates supporting our cells possessed a vibrational mode that dominated the frequency region where both water and TMV absorb.\cite{Liang1998} Thus we turned to study nanospectroscopy of graphene-encapsulated viruses by introducing SiO$_2$ as a substrate with no intrinsic absorptions in the spectral range of interest. Figure \ref{ftir}a shows the AFM topography of a single virion in a graphene liquid cell on a SiO$_2$ substrate, where there exists an additional bottom graphene layer in the cell (see Materials and Methods). We performed a nano-FTIR line scan across the TMV, shown by the green dotted line in Figure \ref{ftir}a, acquiring a spectrum at each pixel in 20 nm steps. The broadband laser source for spectroscopy (see Materials and Methods) was set to the frequency range 1400-1800 cm$^{-1}$, where we expect absorption from the amide I (1660 cm$^{-1}$) and II (1550 cm$^{-1}$) bands in TMV,\cite{Amenabar2013} as well as from water.\cite{Bertie1989}
 
Nano-FTIR absorption spectra for a liquid cell containing TMV are shown in Figs. \ref{ftir}b and \ref{ftir}c, with the width of the virion indicated by the green dashed box in Figure \ref{ftir}c and verified by AFM topography (not shown). The near-field absorption spectrum is approximated by the imaginary part of the scattered \textit{s}-SNOM signal, $Im$[s$_n$e$^{i\phi_n}$] = s$_n$ sin $\phi_n$, which has been shown to correlate well with bulk far-field absorption spectra for weak resonances.\cite{Huth2012} Data in Figure \ref{ftir}b and \ref{ftir}c are normalized to a spectrally flat gold reference. The blue curve in Figure \ref{ftir}d represents a typical nano-FTIR spectrum of the local environment around an isolated virion, averaged over absorption spectra for the 2 $\mu$m line scan shown in Figure \ref{ftir}b. An important feature of the liquid cell nano-FTIR absorption spectra is a minimum in the data near 1610 cm$^{-1}$, whose origin we discuss below. To clearly distinguish the response of the TMV, we plot the background-corrected nano-FTIR absorption, subtracting the average response from the surrounding liquid cell in the linescan shown in Figure \ref{ftir}c. The black curve in Figure \ref{ftir}d shows the resulting nano-FTIR spectrum for TMV, averaged over the green boxed region in Figure \ref{ftir}c. We observe two small spectral features at 1520 cm$^{-1}$ and 1660 cm$^{-1}$ coinciding with the expected amide resonances from the virus proteins. Line scan data shown in Figure\ref{ftir}c are plotted on a logarithmic color scale to emphasize the spectral peaks from the TMV. 

\section{Discussion}
In order to elucidate the nanoimaging and spectroscopy results, it is instructive to more closely examine the microscopic tip-sample interaction specific to our cell and specimens. Contrast in \textit{s}-SNOM images results from the dielectric properties of a small volume underneath the AFM tip. The evanescent fields of the probing tip can penetrate well beneath the sample surface. In practice, the depth of \textit{s}-SNOM sensitivity is effectively several tens of nanometers.\cite{Keilmann2004,Zhang2012,Govyadinov2014} This means that in the case of our liquid cell the tip-sample interaction involves one or two graphene layers, biological matter in water, and the substrate. Thus, there are many sources for forming infrared near-field contrast and also contrast variations due to lateral as well as vertical inhomogeneities in individual layers, including variable amounts of trapped water. Further, unintentional doping of graphene can be nonuniform throughout a single large-area CVD sheet.\cite{Fei2013} During the growth process graphene becomes doped from atmospheric adsorbates or PMMA residue. Although intrinsically IR-transparent, free carriers in graphene interact with phonons in the underlying SiO$_2$ substrate by plasmon-phonon coupling,\cite{Fei2011} leading to enhancement and broadening of the near-field response even far away from the optical phonon of SiO$_2$ at 1128 cm$^{-1}$.

\begin{figure*}
\centering
\includegraphics[hiresbb=true,width=16cm]{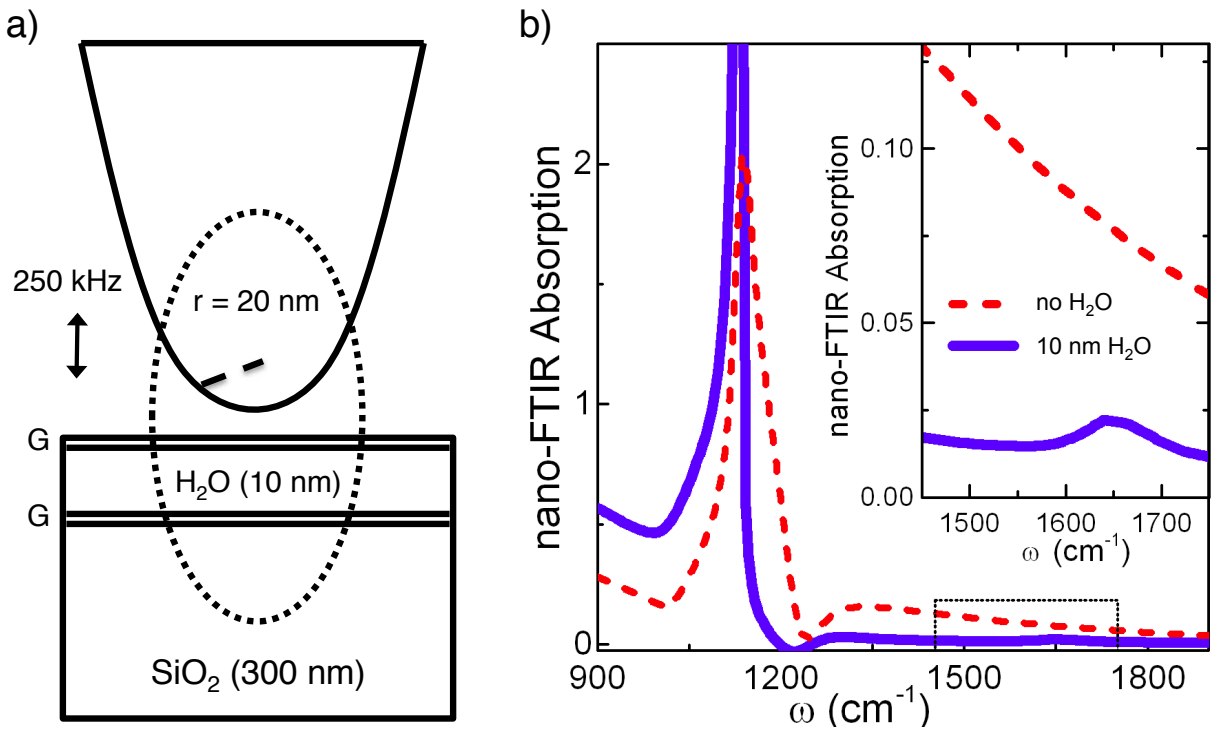}
\caption{Simulation of the near-field response of a liquid cell: (a) schematic showing the lower end of the probing tip, to scale with a 10 nm water layer between two moderately doped monolayer graphene sheets on a SiO$_2$ substrate; (b) resulting near-field absorption spectra with (blue solid line) and without water (red dashed line). The inset shows a 20x expanded view of our experimental spectral range. Clearly, the presence of water in the cell should significantly modify the near-field response.}
\label{fit}
\end{figure*}

Experimental nano-FTIR spectra in Figure \ref{ftir}b of the two-layer graphene liquid cell on SiO$_2$ exhibit a prominent minimum near 1610 cm$^{-1}$. Bulk liquid water possesses a ca. 80 cm$^{-1}$ wide (fwhm) vibrational absorption at 1640 cm$^{-1}$ due to the H-O-H bending mode.\cite{Bertie1989} To simulate the \textit{s}-SNOM response of the multilayer cell structure (Figure \ref{fit}a), we employed the recently developed lightning rod model:\cite{McLeod2014} a quantitative model of the near-field interaction that goes beyond the simple dipole approximation of the tip-sample system,\cite{Keilmann2004} incorporating a realistic probe geometry as well as electrodynamic effects (see Materials and Methods). 

Figure \ref{fit}b gives the predicted \textit{s}-SNOM absorption spectrum without (red dashed curve) and with a 10 nm water layer (blue curve), as previous work has indicated that adequate adhesion from the top graphene layer likely requires less than 15-20 nm of water.\cite{Wood2013} The strong graphene enhancement of the SiO$_2$ surface phonon polariton near $\omega$ = 1100 cm$^{-1}$ results in a large background in the \textit{s}-SNOM response,\cite{Fei2011} which is significantly modified by the presence of water. The 20x expanded curves in the Figure \ref{fit}b inset highlight the spectral range of our nano-FTIR measurements.  

In comparison to what we observe experimentally (blue curve in Figure \ref{ftir}d), we note a deviation of the spectral position of the minimum. However, this is not surprising considering that bulk liquid H$_2$O optical constants were used in the model, which may not accurately reflect possible modifications to the infrared response of graphene-encapsulated water, specifically the energy and line shape of the bending mode of H$_2$O. Nanoconfined or interfacial water is anticipated to have different properties than bulk water\cite{Gorshunov2013,Mante2014,Sechler2010,Algara-Siller2015} and will be the subject of ongoing IR near-field studies. Nonetheless, despite interference from the substrate, we are able to verify the presence of a water layer in the graphene cell. Furthermore, our sensitivity to such a small amount of trapped water among other absorbing materials suggests that more detailed studies on biomacromolecules of sizes < 20 nm are possible using a similar graphene liquid cell architecture. 

Beyond the absorption of water, it is promising that the differential spectrum in Figure \ref{ftir}d (black curve) clearly demonstrates that protein vibrations of a single TMV can be spectroscopically recorded through the graphene layer. The nano-FTIR absorption data for TMV in Figure \ref{ftir}d clearly show two weak resonances at 1520 cm$^{-1}$ and 1660 cm$^{-1}$, indicative of the amide II and I bands,\cite{Brehm2006,Amenabar2013} respectively, superimposed on the broad \textit{s}-SNOM response from the graphene-water-SiO$_2$ subsystem. While the frequency position of the amide I band agrees very well with what has been previously observed in nano-FTIR,\cite{Amenabar2013} the amide II absorption is red-shifted from the expected resonance near 1550 cm$^{-1}$. At these lower energies there is a stronger \textit{s}-SNOM contribution from the SiO$_2$ phonon and doped graphene, which could lead to spectral shifts of the much weaker absorption mode in the TMV. Lastly, in such a nanoconfined environment with a small amount of water, there is considerable hydrostatic pressure imposed by the graphene adhesion to the substrate. High enough local pressure would deform the viral capsid, and might in this way change the vibrational spectrum and possibly the conformations of the proteins.\cite{Wood2013} The large hydrostatic pressure from the graphene confinement can be tuned, by controlling the amount of excess water driven out during sample fabrication (see Supporting Information), to achieve more realistic biological environments in future structures.      

Mica substrates were used for nano-IR imaging experiments because of plasmonic interference effects observed in doped graphene on SiO$_2$ with a CO$_2$ laser.\cite{Fei2012} Conversely, sandwiched TMV samples on SiO$_2$ were more suitable for nanospectroscopy measurements due to the lack of vibrational absorptions associated with the substrate in the frequency range of interest. To realize the full potential of \textit{s}-SNOM imaging and spectroscopy for quantitative characterization of biological materials in aqueous media, the liquid cell architecture can be optimized by selecting spectrally flat substrates such as Si or Au\cite{Mastel2015} or by using free-standing graphene nanosandwich cells transferred to TEM grids.\cite{Wang2014} Additionally, future iterations of such graphene-based liquid cells could employ other isotopes of water to shift the bending mode vibration out of the protein frequency region.\cite{Bertie1989} Other 2D materials besides graphene can also be explored for use as a nanofluidic lid enabling \textit{s}-SNOM investigation, such as BN,\cite{Zhou2013} transition metal dichalcogenides (TMDs),\cite{Jariwala2014} or phosphorene.\cite{Li2014,Wood2014} BN might be an especially attractive alternative by virtue of its hyperbolic dispersion and ability for subdiffractional focusing in the mid-infrared.\cite{Dai2015} Finally, while we have demonstrated the general concept of near-field nano-imaging and spectroscopy through an ultrathin barrier, the underlying liquid cell structure can in principle be more sophisticated, utilizing advanced design principles from the fields of micro- and nanofluidics. 

\section{Conclusion}
We have performed \textit{s}-SNOM on TMV trapped together with water underneath large-area monolayer graphene. We were able to image individual virions through this graphene "lid", registering significant contrast in both amplitude and phase of the scattered near-field signal. We observed in the near-field response spectral features associated with trapped water, superimposed on a background contribution from the underlying SiO$_2$ substrate. Further we measured, on a single virion inside the graphene liquid cell, an absorption spectrum due to the amide I and II vibrational bands of virus proteins. Therefore, our nanospectroscopic probe was able to register nanometer quantities of water, suggesting studies of other macromolecular materials using a similar sample architecture. We have demonstrated that scattering-based near-field techniques are applicable to biological samples in aqueous media. Our work sets the stage for further nano-FTIR studies of biological and molecular materials, utilizing optimized liquid cells employing ultrathin lids made of graphene and other 2D nanomaterials.

\section{Materials and Methods}
\subsection{Graphene Liquid Cell Sample Preparation}
Large-area graphene is grown from chemical vapor deposition (CVD) atop Cu foils. A lower methane-to-hydrogen ratio is used to give higher monolayer coverage. The method for trapping water with viruses beneath graphene is a wet transfer similar to what has been established previously for water alone.\cite{He2012,Wood2013} The graphene is transferred with poly(methyl methacrylate) (PMMA) onto either mica or SiO$_2$/Si substrates. Successive deionized water baths clean the graphene films from residual etchant contamination after the transfer process. There is some remaining water from the wet transfer process that is trapped between the graphene overlayer and the underlying substrate surface. The PMMA transfer scaffold is dissolved by overnight soaking in chloroform. For SiO$_2$/Si substrates, an additional bottom layer of graphene was used to prevent water from escaping through the oxide, creating a graphene "nano-sandwich." Both mica and SiO$_2$-based substrates function as suitable \textit{s}-SNOM-compatible liquid cells employing a graphene lid, as shown in Figure \ref{schem}b. Further details about graphene liquid cell sample preparation can be found in the Supporting Information.

\subsection{\textit{s}-SNOM Measurements}
For \textit{s}-SNOM imaging and nano-FTIR measurements, we used a commercial near-field microscope (Neaspec GmbH) based on an AFM operating in tapping mode at a frequency of $\sim$ 250 kHz, with PtIr-coated cantilevers of nominal tip radius 20-30 nm (Arrow NCPt, NanoWorld AG) (Figure \ref{schem}a). The single frequency IR nanoimaging results were obtained by illuminating the tip with a CO$_2$ laser beam (Access Laser Co.). The scattered light from the tip is focused onto a HgCdTe detector (Kolmar Technologies). The \textit{s}-SNOM detector signal comprises a far-field background and the local near-field interaction, the latter being modulated by the oscillating AFM tip. The far-field background is strongly suppressed by demodulating the \textit{s}-SNOM signal at harmonics of the tip tapping frequency.\cite{Keilmann2004} Pseudoheterodyne detection\cite{Ocelic2006} with the reference mirror oscillating at 300 Hz is used to recover both amplitude s$_n$ and phase $\phi_n$, where n is the harmonic order, typically 2 or 3.

For nano-FTIR experiments, we used a broadband coherent source (Lasnix, Germany) based on difference-frequency generation.\cite{Keilmann2012,Amarie2009} The mid-infrared beam is generated by the nonlinear mixing of a train of near-IR Er-based fiber laser pulses with synchronous frequency-offset IR supercontinuum pulses (Toptica Photonics, Germany) inside a GaSe crystal.\cite{Gambetta2008} The generated mid-IR spectrum spans 250-300 cm$^{-1}$ and is tunable between 650 and 2200 cm$^{-1}$. The asymmetric Michelson interferometer (Figure \ref{schem}a) was used to record the \textit{s}-SNOM signal, by scanning the reference mirror over up to 1500 $\mu$m travel range.\cite{Amarie2009} Nano-FTIR spectra are typically collected in 60 s at 8 cm$^{-1}$ resolution, and normalized to those from a Au reference. Nano-FTIR absorption is computed as $Im$[s$_n$e$^{i\phi_n}$] = s$_n$ sin $\phi_n$.\cite{Huth2012} 

\subsection{\textit{s}-SNOM Response Simulations}
Near-field scattering amplitudes were calculated using the lightning rod model.\cite{McLeod2014} The model (Figure \ref{fit}a) assumes a hyperboloid to approximate the conical structure of the AFM tip with apex radius r = 20 nm and taper angle $\sim$ 20$^{\circ}$. For the simulations, we used tabulated H$_2$O and SiO$_2$ optical constants taken from the literature.\cite{Hale1973,Herzinger1998} For the top and bottom graphene layers, we assume a moderate doping level denoted by a chemical potential, $\mu$ = 2000 cm$^{-1}$, and apply a form for the optical conductivity calculated within the random phase approximation.\cite{Fei2011}

\section{Acknowledgements}
Research at UCSD is supported by ONR. Development of nanospectroscopy instrumentation at UCSD is supported by DOE-BES. D. N. B. is the Moore Investigator in Quantum Materials Grant GBMF-4533. The authors would like to acknowledge E. A. Carrion for providing graphene, J.-W. Do for assistance during graphene annealing, and Y. Chen for AFM assistance during sample preparation.

The authors declare the following competing financial interest(s): F. Keilmann is co-founder of Neaspec and Lasnix, producers of the s-SNOM and infrared source used in this study. The remaining authors declare no competing financial interests.


\bibliography{bib}

\providecommand{\latin}[1]{#1}
\providecommand*\mcitethebibliography{\thebibliography}
\csname @ifundefined\endcsname{endmcitethebibliography}
  {\let\endmcitethebibliography\endthebibliography}{}
\begin{mcitethebibliography}{68}
\providecommand*\natexlab[1]{#1}
\providecommand*\mciteSetBstSublistMode[1]{}
\providecommand*\mciteSetBstMaxWidthForm[2]{}
\providecommand*\mciteBstWouldAddEndPuncttrue
  {\def\EndOfBibitem{\unskip.}}
\providecommand*\mciteBstWouldAddEndPunctfalse
  {\let\EndOfBibitem\relax}
\providecommand*\mciteSetBstMidEndSepPunct[3]{}
\providecommand*\mciteSetBstSublistLabelBeginEnd[3]{}
\providecommand*\EndOfBibitem{}
\mciteSetBstSublistMode{f}
\mciteSetBstMaxWidthForm{subitem}{(\alph{mcitesubitemcount})}
\mciteSetBstSublistLabelBeginEnd
  {\mcitemaxwidthsubitemform\space}
  {\relax}
  {\relax}

\bibitem[Movasaghi \latin{et~al.}(2008)Movasaghi, Rehman, and
  ur~Rehman]{Movasaghi2008}
Movasaghi,~Z.; Rehman,~S.; ur~Rehman,~D.~I. Fourier Transform Infrared (FTIR)
  Spectroscopy of Biological Tissues. \emph{Appl. Spectrosc. Rev.}
  \textbf{2008}, \emph{43}, 134--179\relax
\mciteBstWouldAddEndPuncttrue
\mciteSetBstMidEndSepPunct{\mcitedefaultmidpunct}
{\mcitedefaultendpunct}{\mcitedefaultseppunct}\relax
\EndOfBibitem
\bibitem[Baker \latin{et~al.}(2014)Baker, Trevisan, Bassan, Bhargava, Butler,
  Dorling, Fielden, Fogarty, Fullwood, Heys, Hughes, Lasch, Martin-Hirsch,
  Obinaju, Sockalingum, Sulé-Suso, Strong, Walsh, Wood, Gardner, and
  Martin]{Baker2014}
Baker,~M.~J.; Trevisan,~J.; Bassan,~P.; Bhargava,~R.; Butler,~H.~J.;
  Dorling,~K.~M.; Fielden,~P.~R.; Fogarty,~S.~W.; Fullwood,~N.~J.; Heys,~K.~A.
  \latin{et~al.}  Using Fourier Transform IR Spectroscopy to Analyze Biological
  Materials. \emph{Nat. Protoc.} \textbf{2014}, \emph{9}, 1771--1791\relax
\mciteBstWouldAddEndPuncttrue
\mciteSetBstMidEndSepPunct{\mcitedefaultmidpunct}
{\mcitedefaultendpunct}{\mcitedefaultseppunct}\relax
\EndOfBibitem
\bibitem[Juszczyk \latin{et~al.}(2009)Juszczyk, Kolodziejczyk, and
  Grzonka]{Juszczyk2009}
Juszczyk,~P.; Kolodziejczyk,~A.~S.; Grzonka,~Z. FTIR Spectroscopic Studies on
  Aggregation Process of the $\beta$-Amyloid 11-28 Fragment and its Variants.
  \emph{J. Peptide Sci.} \textbf{2009}, \emph{15}, 23--29\relax
\mciteBstWouldAddEndPuncttrue
\mciteSetBstMidEndSepPunct{\mcitedefaultmidpunct}
{\mcitedefaultendpunct}{\mcitedefaultseppunct}\relax
\EndOfBibitem
\bibitem[Ami \latin{et~al.}(2013)Ami, Mereghetti, and Maria~Doglia]{Ami2013}
Ami,~D.; Mereghetti,~P.; Maria~Doglia,~S. In \emph{Multivariate Analysis in
  Management, Engineering and the Sciences}; Freitas,~L., Ed.; InTech, 2013;
  Chapter 10, pp 189--220\relax
\mciteBstWouldAddEndPuncttrue
\mciteSetBstMidEndSepPunct{\mcitedefaultmidpunct}
{\mcitedefaultendpunct}{\mcitedefaultseppunct}\relax
\EndOfBibitem
\bibitem[Gucciardi(2009)]{Gucciardi2009}
Gucciardi,~P.~G. In \emph{Applied Scanning Probe Methods XII}; Bhushan,~B.,
  Fuchs,~H., Eds.; Springer Berlin Heidelberg, 2009; Chapter 11, pp
  49--68\relax
\mciteBstWouldAddEndPuncttrue
\mciteSetBstMidEndSepPunct{\mcitedefaultmidpunct}
{\mcitedefaultendpunct}{\mcitedefaultseppunct}\relax
\EndOfBibitem
\bibitem[Haris and Chapman(1994)Haris, and Chapman]{Haris1994}
Haris,~P.~I.; Chapman,~D. In \emph{Microscopy, Optical Spectroscopy, and
  Macroscopic Techniques}, 1st ed.; Jones,~C., Mulloy,~B., Thomas,~A.~H., Eds.;
  Methods in Molecular Biology; Humana Press, 1994; Vol.~22; Chapter 14, pp
  183--202\relax
\mciteBstWouldAddEndPuncttrue
\mciteSetBstMidEndSepPunct{\mcitedefaultmidpunct}
{\mcitedefaultendpunct}{\mcitedefaultseppunct}\relax
\EndOfBibitem
\bibitem[Hillenbrand and Keilmann(2002)Hillenbrand, and
  Keilmann]{Hillenbrand2002a}
Hillenbrand,~R.; Keilmann,~F. Material-Specific Mapping of
  Metal/Semiconductor/Dielectric Nanosystems at 10 nm Resolution by
  Backscattering Near-Field Optical Microscopy. \emph{Appl. Phys. Lett.}
  \textbf{2002}, \emph{80}, 25--27\relax
\mciteBstWouldAddEndPuncttrue
\mciteSetBstMidEndSepPunct{\mcitedefaultmidpunct}
{\mcitedefaultendpunct}{\mcitedefaultseppunct}\relax
\EndOfBibitem
\bibitem[Hillenbrand \latin{et~al.}(2001)Hillenbrand, Knoll, and
  Keilmann]{Hillenbrand2001}
Hillenbrand,~R.; Knoll,~B.; Keilmann,~F. Pure Optical Contrast in
  Scattering-Type Scanning Near-Field Microscopy. \emph{J. Microsc.}
  \textbf{2001}, \emph{202}, 77--83\relax
\mciteBstWouldAddEndPuncttrue
\mciteSetBstMidEndSepPunct{\mcitedefaultmidpunct}
{\mcitedefaultendpunct}{\mcitedefaultseppunct}\relax
\EndOfBibitem
\bibitem[Keilmann and Hillenbrand(2004)Keilmann, and Hillenbrand]{Keilmann2004}
Keilmann,~F.; Hillenbrand,~R. Near-Field Microscopy by Elastic Light Scattering
  from a Tip. \emph{Philos. Trans. R. Soc., A} \textbf{2004}, \emph{362},
  787--805\relax
\mciteBstWouldAddEndPuncttrue
\mciteSetBstMidEndSepPunct{\mcitedefaultmidpunct}
{\mcitedefaultendpunct}{\mcitedefaultseppunct}\relax
\EndOfBibitem
\bibitem[Atkin \latin{et~al.}(2012)Atkin, Berweger, Jones, and
  Raschke]{Atkin2012}
Atkin,~J.~M.; Berweger,~S.; Jones,~A.~C.; Raschke,~M.~B. Nano-Optical Imaging
  and Spectroscopy of Order, Phases, and Domains in Complex Solids. \emph{Adv.
  Phys.} \textbf{2012}, \emph{61}, 745--842\relax
\mciteBstWouldAddEndPuncttrue
\mciteSetBstMidEndSepPunct{\mcitedefaultmidpunct}
{\mcitedefaultendpunct}{\mcitedefaultseppunct}\relax
\EndOfBibitem
\bibitem[Keilmann and Amarie(2012)Keilmann, and Amarie]{Keilmann2012}
Keilmann,~F.; Amarie,~S. Mid-infrared Frequency Comb Spanning an Octave Based
  on an Er Fiber Laser and Difference-Frequency Generation. \emph{J. Infrared,
  Millimeter, Terahertz Waves} \textbf{2012}, \emph{33}, 479--484\relax
\mciteBstWouldAddEndPuncttrue
\mciteSetBstMidEndSepPunct{\mcitedefaultmidpunct}
{\mcitedefaultendpunct}{\mcitedefaultseppunct}\relax
\EndOfBibitem
\bibitem[Amarie \latin{et~al.}(2009)Amarie, Ganz, and Keilmann]{Amarie2009}
Amarie,~S.; Ganz,~T.; Keilmann,~F. Mid-Infrared Near-Field Spectroscopy.
  \emph{Opt. Express} \textbf{2009}, \emph{17}, 21794--21801\relax
\mciteBstWouldAddEndPuncttrue
\mciteSetBstMidEndSepPunct{\mcitedefaultmidpunct}
{\mcitedefaultendpunct}{\mcitedefaultseppunct}\relax
\EndOfBibitem
\bibitem[Huth \latin{et~al.}(2011)Huth, Schnell, Wittborn, Ocelic, and
  Hillenbrand]{Huth2011}
Huth,~F.; Schnell,~M.; Wittborn,~J.; Ocelic,~N.; Hillenbrand,~R.
  Infrared-Spectroscopic Nanoimaging with a Thermal Source. \emph{Nat. Mater.}
  \textbf{2011}, \emph{10}, 352--356\relax
\mciteBstWouldAddEndPuncttrue
\mciteSetBstMidEndSepPunct{\mcitedefaultmidpunct}
{\mcitedefaultendpunct}{\mcitedefaultseppunct}\relax
\EndOfBibitem
\bibitem[Dai \latin{et~al.}(2014)Dai, Fei, Ma, Rodin, Wagner, McLeod, Liu,
  Gannett, Regan, Watanabe, Taniguchi, Thiemens, Dominguez, Neto, Zettl,
  Keilmann, Jarillo-Herrero, Fogler, and Basov]{Dai2014}
Dai,~S.; Fei,~Z.; Ma,~Q.; Rodin,~A.~S.; Wagner,~M.; McLeod,~A.~S.; Liu,~M.~K.;
  Gannett,~W.; Regan,~W.; Watanabe,~K. \latin{et~al.}  Tunable Phonon
  Polaritons in Atomically Thin van der Waals Crystals of Boron Nitride.
  \emph{Science} \textbf{2014}, \emph{343}, 1125--1129\relax
\mciteBstWouldAddEndPuncttrue
\mciteSetBstMidEndSepPunct{\mcitedefaultmidpunct}
{\mcitedefaultendpunct}{\mcitedefaultseppunct}\relax
\EndOfBibitem
\bibitem[Schnell \latin{et~al.}(2014)Schnell, Carney, and
  Hillenbrand]{Schnell2014}
Schnell,~M.; Carney,~P.~S.; Hillenbrand,~R. Synthetic Optical Holography for
  Rapid Nanoimaging. \emph{Nat. Commun.} \textbf{2014}, \emph{5},
  3499--3508\relax
\mciteBstWouldAddEndPuncttrue
\mciteSetBstMidEndSepPunct{\mcitedefaultmidpunct}
{\mcitedefaultendpunct}{\mcitedefaultseppunct}\relax
\EndOfBibitem
\bibitem[Gucciardi \latin{et~al.}(2008)Gucciardi, Bachelier, Stranick, and
  Allegrini]{Gucciardi2008}
Gucciardi,~P.~G.; Bachelier,~G.; Stranick,~S.~J.; Allegrini,~M. In
  \emph{Applied Scanning Probe Methods VIII}; Bhushan,~B., Fuchs,~H.,
  Tomitori,~M., Eds.; Nano Science and Technolgy; Springer-Verlag Berlin
  Heidelberg, 2008; Chapter 1, pp 1--29\relax
\mciteBstWouldAddEndPuncttrue
\mciteSetBstMidEndSepPunct{\mcitedefaultmidpunct}
{\mcitedefaultendpunct}{\mcitedefaultseppunct}\relax
\EndOfBibitem
\bibitem[Brehm \latin{et~al.}(2006)Brehm, Taubner, Hillenbrand, and
  Keilmann]{Brehm2006}
Brehm,~M.; Taubner,~T.; Hillenbrand,~R.; Keilmann,~F. Infrared Spectroscopic
  Mapping of Single Nanoparticles and Viruses at Nanoscale Resolution.
  \emph{Nano Lett.} \textbf{2006}, \emph{6}, 1307--1310\relax
\mciteBstWouldAddEndPuncttrue
\mciteSetBstMidEndSepPunct{\mcitedefaultmidpunct}
{\mcitedefaultendpunct}{\mcitedefaultseppunct}\relax
\EndOfBibitem
\bibitem[Amenabar \latin{et~al.}(2013)Amenabar, Poly, Nuansing, Hubrich,
  Govyadinov, Huth, Krutokhvostov, Zhang, Knez, Heberle, Bittner, and
  Hillenbrand]{Amenabar2013}
Amenabar,~I.; Poly,~S.; Nuansing,~W.; Hubrich,~E.~H.; Govyadinov,~A.~A.;
  Huth,~F.; Krutokhvostov,~R.; Zhang,~L.; Knez,~M.; Heberle,~J. \latin{et~al.}
  Structural Analysis and Mapping of Individual Protein Complexes by Infrared
  Nanospectroscopy. \emph{Nat. Commun.} \textbf{2013}, \emph{4}, 1--9\relax
\mciteBstWouldAddEndPuncttrue
\mciteSetBstMidEndSepPunct{\mcitedefaultmidpunct}
{\mcitedefaultendpunct}{\mcitedefaultseppunct}\relax
\EndOfBibitem
\bibitem[Berweger \latin{et~al.}(2013)Berweger, Nguyen, Muller, Bechtel,
  Perkins, and Raschke]{Berweger2013}
Berweger,~S.; Nguyen,~D.~M.; Muller,~E.~A.; Bechtel,~H.~A.; Perkins,~T.~T.;
  Raschke,~M.~B. Nano-Chemical Infrared Imaging of Membrane Proteins in Lipid
  Bilayers. \emph{J. Am. Chem. Soc.} \textbf{2013}, \emph{135},
  18292--18295\relax
\mciteBstWouldAddEndPuncttrue
\mciteSetBstMidEndSepPunct{\mcitedefaultmidpunct}
{\mcitedefaultendpunct}{\mcitedefaultseppunct}\relax
\EndOfBibitem
\bibitem[Legleiter and Kowalewski(2005)Legleiter, and
  Kowalewski]{Legleiter2005}
Legleiter,~J.; Kowalewski,~T. Insights into Fluid Tapping-Mode Atomic Force
  Microscopy Provided by Numerical Simulations. \emph{Appl. Phys. Lett.}
  \textbf{2005}, \emph{87}, 163120\relax
\mciteBstWouldAddEndPuncttrue
\mciteSetBstMidEndSepPunct{\mcitedefaultmidpunct}
{\mcitedefaultendpunct}{\mcitedefaultseppunct}\relax
\EndOfBibitem
\bibitem[Preiner \latin{et~al.}(2007)Preiner, Tang, Pastushenko, and
  Hinterdorfer]{Preiner2007}
Preiner,~J.; Tang,~J.; Pastushenko,~V.; Hinterdorfer,~P. Higher Harmonic Atomic
  Force Microscopy: Imaging of Biological Membranes in Liquid. \emph{Phys. Rev.
  Lett.} \textbf{2007}, \emph{99}, 046102\relax
\mciteBstWouldAddEndPuncttrue
\mciteSetBstMidEndSepPunct{\mcitedefaultmidpunct}
{\mcitedefaultendpunct}{\mcitedefaultseppunct}\relax
\EndOfBibitem
\bibitem[Xu \latin{et~al.}(2010)Xu, Cao, and Heath]{Xu2010}
Xu,~K.; Cao,~P.; Heath,~J.~R. Graphene Visualizes the First Water Adlayers on
  Mica at Ambient Conditions. \emph{Science} \textbf{2010}, \emph{329},
  1188--1191\relax
\mciteBstWouldAddEndPuncttrue
\mciteSetBstMidEndSepPunct{\mcitedefaultmidpunct}
{\mcitedefaultendpunct}{\mcitedefaultseppunct}\relax
\EndOfBibitem
\bibitem[He \latin{et~al.}(2012)He, Wood, Doidge, Pop, and Lyding]{He2012}
He,~K.~T.; Wood,~J.~D.; Doidge,~G.~P.; Pop,~E.; Lyding,~J.~W. Scanning
  Tunneling Microscopy Study and Nanomanipulation of Graphene-Coated Water on
  Mica. \emph{Nano Lett.} \textbf{2012}, \emph{12}, 2665--2672\relax
\mciteBstWouldAddEndPuncttrue
\mciteSetBstMidEndSepPunct{\mcitedefaultmidpunct}
{\mcitedefaultendpunct}{\mcitedefaultseppunct}\relax
\EndOfBibitem
\bibitem[Wood \latin{et~al.}(2012)Wood, Schmucker, Haasch, Doidge, Nienhaus,
  Damhorst, Lyons, Gruebele, Bashir, Pop, and Lyding]{Wood2012}
Wood,~J.~D.; Schmucker,~S.~W.; Haasch,~R.~T.; Doidge,~G.~P.; Nienhaus,~L.;
  Damhorst,~G.~L.; Lyons,~A.~S.; Gruebele,~M.; Bashir,~R.; Pop,~E.
  \latin{et~al.}  Improved Graphene Growth and Fluorination on Cu with Clean
  Transfer to Surfaces. IEEE Conference on Nanotechnology (IEEE-NANO). 2012; pp
  1--4\relax
\mciteBstWouldAddEndPuncttrue
\mciteSetBstMidEndSepPunct{\mcitedefaultmidpunct}
{\mcitedefaultendpunct}{\mcitedefaultseppunct}\relax
\EndOfBibitem
\bibitem[Novoselov \latin{et~al.}(2012)Novoselov, Fal'ko, Colombo, Gellert,
  Schwab, and Kim]{Novoselov2012}
Novoselov,~K.~S.; Fal'ko,~V.~I.; Colombo,~L.; Gellert,~P.~R.; Schwab,~M.~G.;
  Kim,~K. A Roadmap for Graphene. \emph{Nature} \textbf{2012}, \emph{490},
  192--200\relax
\mciteBstWouldAddEndPuncttrue
\mciteSetBstMidEndSepPunct{\mcitedefaultmidpunct}
{\mcitedefaultendpunct}{\mcitedefaultseppunct}\relax
\EndOfBibitem
\bibitem[Basov \latin{et~al.}(2014)Basov, Fogler, Lanzara, Wang, and
  Zhang]{Basov2014}
Basov,~D.~N.; Fogler,~M.; Lanzara,~A.; Wang,~F.; Zhang,~Y. \textit{Colloquium}
  : Graphene Spectroscopy. \emph{Rev. Mod. Phys.} \textbf{2014}, \emph{86},
  959--994\relax
\mciteBstWouldAddEndPuncttrue
\mciteSetBstMidEndSepPunct{\mcitedefaultmidpunct}
{\mcitedefaultendpunct}{\mcitedefaultseppunct}\relax
\EndOfBibitem
\bibitem[Wang \latin{et~al.}(2014)Wang, Qiao, Shokuhfar, and Klie]{Wang2014}
Wang,~C.; Qiao,~Q.; Shokuhfar,~T.; Klie,~R.~F. High-Resolution Electron
  Microscopy and Spectroscopy of Ferritin in Biocompatible Graphene Liquid
  Cells and Graphene Sandwiches. \emph{Adv. Mater.} \textbf{2014}, \emph{26},
  3410--3414\relax
\mciteBstWouldAddEndPuncttrue
\mciteSetBstMidEndSepPunct{\mcitedefaultmidpunct}
{\mcitedefaultendpunct}{\mcitedefaultseppunct}\relax
\EndOfBibitem
\bibitem[Wood(2013)]{Wood2013}
Wood,~J.~D. Large-Scale Growth, Fluorination, Clean Transfer, and Layering of
  Graphene and Related Nanomaterials. Ph.D.\ thesis, University of Illinois at
  Urbana-Champaign, 2013\relax
\mciteBstWouldAddEndPuncttrue
\mciteSetBstMidEndSepPunct{\mcitedefaultmidpunct}
{\mcitedefaultendpunct}{\mcitedefaultseppunct}\relax
\EndOfBibitem
\bibitem[Zhong \latin{et~al.}(2009)Zhong, Zeng, Lu, Wang, Gong, Yan, Huang, and
  Chen]{Zhong2009}
Zhong,~L.; Zeng,~G.; Lu,~X.; Wang,~R.~C.; Gong,~G.; Yan,~L.; Huang,~D.;
  Chen,~Z.~W. NSOM/QD-Based Direct Visualization of CD3-Induced and
  CD28-Enhanced Nanospatial Coclustering of TCR and Coreceptor in Nanodomains
  in T Cell Activation. \emph{PLoS ONE} \textbf{2009}, \emph{4}, e5945--\relax
\mciteBstWouldAddEndPuncttrue
\mciteSetBstMidEndSepPunct{\mcitedefaultmidpunct}
{\mcitedefaultendpunct}{\mcitedefaultseppunct}\relax
\EndOfBibitem
\bibitem[Hu \latin{et~al.}(2009)Hu, Chen, Wang, Wang, Ma, Cai, Chen, and
  Chen]{Hu2009}
Hu,~M.; Chen,~J.; Wang,~J.; Wang,~X.; Ma,~S.; Cai,~J.; Chen,~C.~Y.; Chen,~Z.~W.
  AFM- and NSOM-based Force Spectroscopy and Distribution Analysis of CD69
  Molecules on Human CD4+ T Cell Membrane. \emph{J. Mol. Recognit.}
  \textbf{2009}, \emph{22}, 516--520\relax
\mciteBstWouldAddEndPuncttrue
\mciteSetBstMidEndSepPunct{\mcitedefaultmidpunct}
{\mcitedefaultendpunct}{\mcitedefaultseppunct}\relax
\EndOfBibitem
\bibitem[Dickenson \latin{et~al.}(2010)Dickenson, Armendariz, Huckabay,
  Livanec, and Dunn]{Dickenson2010}
Dickenson,~N.~E.; Armendariz,~K.~P.; Huckabay,~H.~A.; Livanec,~P.~W.;
  Dunn,~R.~C. Near-Field Scanning Optical Microscopy: A tool for Nanometric
  Exploration of Biological Membranes. \emph{Anal. Bioanal. Chem.}
  \textbf{2010}, \emph{396}, 31--43\relax
\mciteBstWouldAddEndPuncttrue
\mciteSetBstMidEndSepPunct{\mcitedefaultmidpunct}
{\mcitedefaultendpunct}{\mcitedefaultseppunct}\relax
\EndOfBibitem
\bibitem[Lewis \latin{et~al.}(1999)Lewis, Radko, Ben~Ami, Palanker, and
  Lieberman]{Lewis1999}
Lewis,~A.; Radko,~A.; Ben~Ami,~N.; Palanker,~D.; Lieberman,~K. Near-Field
  Scanning Optical Microscopy in Cell Biology. \emph{Trends Cell Biol.}
  \textbf{1999}, \emph{9}, 70--73\relax
\mciteBstWouldAddEndPuncttrue
\mciteSetBstMidEndSepPunct{\mcitedefaultmidpunct}
{\mcitedefaultendpunct}{\mcitedefaultseppunct}\relax
\EndOfBibitem
\bibitem[Edidin(2001)]{Edidin2001}
Edidin,~M. Near-Field Scanning Optical Microscopy, a Siren Call to Biology.
  \emph{Traffic} \textbf{2001}, \emph{2}, 797--803\relax
\mciteBstWouldAddEndPuncttrue
\mciteSetBstMidEndSepPunct{\mcitedefaultmidpunct}
{\mcitedefaultendpunct}{\mcitedefaultseppunct}\relax
\EndOfBibitem
\bibitem[de~Lange \latin{et~al.}(2001)de~Lange, Cambi, Huijbens, de~Bakker,
  Rensen, Garcia-Parajo, van Hulst, and Figdor]{Lange2001}
de~Lange,~F.; Cambi,~A.; Huijbens,~R.; de~Bakker,~B.; Rensen,~W.;
  Garcia-Parajo,~M.; van Hulst,~N.; Figdor,~C.~G. Cell Biology Beyond the
  Diffraction Limit: Near-Field Scanning Optical microscopy. \emph{J. Cell
  Sci.} \textbf{2001}, \emph{114}, 4153--4160\relax
\mciteBstWouldAddEndPuncttrue
\mciteSetBstMidEndSepPunct{\mcitedefaultmidpunct}
{\mcitedefaultendpunct}{\mcitedefaultseppunct}\relax
\EndOfBibitem
\bibitem[Hinterdorfer \latin{et~al.}(2011)Hinterdorfer, Garcia-Parajo, and
  Dufrêne]{Hinterdorfer2011}
Hinterdorfer,~P.; Garcia-Parajo,~M.~F.; Dufrêne,~Y.~F. Single-Molecule Imaging
  of Cell Surfaces Using Near-Field Nanoscopy. \emph{Acc. Chem. Res.}
  \textbf{2011}, \emph{45}, 327--336\relax
\mciteBstWouldAddEndPuncttrue
\mciteSetBstMidEndSepPunct{\mcitedefaultmidpunct}
{\mcitedefaultendpunct}{\mcitedefaultseppunct}\relax
\EndOfBibitem
\bibitem[Kapkiai \latin{et~al.}(2004)Kapkiai, Moore-Nichols, Carnell,
  Krogmeier, and Dunn]{Kapkiai2004}
Kapkiai,~L.~K.; Moore-Nichols,~D.; Carnell,~J.; Krogmeier,~J.~R.; Dunn,~R.~C.
  Hybrid Near-Field Scanning Optical Microscopy Tips for Live Cell
  Measurements. \emph{Appl. Phys. Lett.} \textbf{2004}, \emph{84},
  3750--3752\relax
\mciteBstWouldAddEndPuncttrue
\mciteSetBstMidEndSepPunct{\mcitedefaultmidpunct}
{\mcitedefaultendpunct}{\mcitedefaultseppunct}\relax
\EndOfBibitem
\bibitem[van Zanten \latin{et~al.}(2010)van Zanten, Cambi, and
  Garcia-Parajo]{Zanten2010}
van Zanten,~T.~S.; Cambi,~A.; Garcia-Parajo,~M.~F. A Nanometer Scale Optical
  View on the Compartmentalization of Cell Membranes. \emph{Biochim. Biophys.
  Acta, Biomembr.} \textbf{2010}, \emph{1798}, 777--787\relax
\mciteBstWouldAddEndPuncttrue
\mciteSetBstMidEndSepPunct{\mcitedefaultmidpunct}
{\mcitedefaultendpunct}{\mcitedefaultseppunct}\relax
\EndOfBibitem
\bibitem[Dunn(1999)]{Dunn1999}
Dunn,~R.~C. Near-Field Scanning Optical Microscopy. \emph{Chem. Rev.}
  \textbf{1999}, \emph{99}, 2891--2928\relax
\mciteBstWouldAddEndPuncttrue
\mciteSetBstMidEndSepPunct{\mcitedefaultmidpunct}
{\mcitedefaultendpunct}{\mcitedefaultseppunct}\relax
\EndOfBibitem
\bibitem[Herrmann \latin{et~al.}(2009)Herrmann, Neuberth, Wissler, Pérez,
  Gradl, and Naber]{Herrmann2009}
Herrmann,~M.; Neuberth,~N.; Wissler,~J.; Pérez,~J.; Gradl,~D.; Naber,~A.
  Near-Field Optical Study of Protein Transport Kinetics at a Single Nuclear
  Pore. \emph{Nano Lett.} \textbf{2009}, \emph{9}, 3330--3336\relax
\mciteBstWouldAddEndPuncttrue
\mciteSetBstMidEndSepPunct{\mcitedefaultmidpunct}
{\mcitedefaultendpunct}{\mcitedefaultseppunct}\relax
\EndOfBibitem
\bibitem[Trinh \latin{et~al.}(2011)Trinh, Odorico, Bellanger, Jacquemond,
  Parot, and Pellequer]{Trinh2011}
Trinh,~M.-H.; Odorico,~M.; Bellanger,~L.; Jacquemond,~M.; Parot,~P.;
  Pellequer,~J.-L. Tobacco Mosaic Virus as an AFM Tip Calibrator. \emph{J. Mol.
  Recognit.} \textbf{2011}, \emph{24}, 503--510\relax
\mciteBstWouldAddEndPuncttrue
\mciteSetBstMidEndSepPunct{\mcitedefaultmidpunct}
{\mcitedefaultendpunct}{\mcitedefaultseppunct}\relax
\EndOfBibitem
\bibitem[Alonso \latin{et~al.}(2013)Alonso, G\'orzny, and Bittner]{Alonso2013}
Alonso,~J.~M.; G\'orzny,~M.~L.; Bittner,~A.~M. The Physics of Tobacco Mosaic
  Virus and Virus-based Devices in Biotechnology. \emph{Trends Biotechnol.}
  \textbf{2013}, \emph{31}, 530--538\relax
\mciteBstWouldAddEndPuncttrue
\mciteSetBstMidEndSepPunct{\mcitedefaultmidpunct}
{\mcitedefaultendpunct}{\mcitedefaultseppunct}\relax
\EndOfBibitem
\bibitem[Fei \latin{et~al.}(2013)Fei, Rodin, Gannett, Dai, Regan, Wagner, Liu,
  McLeod, Dominguez, Thiemens, Castro~Neto, Keilmann, Zettl, Hillenbrand,
  Fogler, and Basov]{Fei2013}
Fei,~Z.; Rodin,~A.~S.; Gannett,~W.; Dai,~S.; Regan,~W.; Wagner,~M.; Liu,~M.~K.;
  McLeod,~A.~S.; Dominguez,~G.; Thiemens,~M. \latin{et~al.}  Electronic and
  Plasmonic Phenomena at Graphene Grain Boundaries. \emph{Nat. Nanotechnol.}
  \textbf{2013}, \emph{8}, 821--825\relax
\mciteBstWouldAddEndPuncttrue
\mciteSetBstMidEndSepPunct{\mcitedefaultmidpunct}
{\mcitedefaultendpunct}{\mcitedefaultseppunct}\relax
\EndOfBibitem
\bibitem[Hillenbrand \latin{et~al.}(2002)Hillenbrand, Taubner, and
  Keilmann]{Hillenbrand2002}
Hillenbrand,~R.; Taubner,~T.; Keilmann,~F. Phonon-enhanced Light-Matter
  Interaction at the Nanometre Scale. \emph{Nature} \textbf{2002}, \emph{418},
  159--162\relax
\mciteBstWouldAddEndPuncttrue
\mciteSetBstMidEndSepPunct{\mcitedefaultmidpunct}
{\mcitedefaultendpunct}{\mcitedefaultseppunct}\relax
\EndOfBibitem
\bibitem[Taubner \latin{et~al.}(2004)Taubner, Hillenbrand, and
  Keilmann]{Taubner2004}
Taubner,~T.; Hillenbrand,~R.; Keilmann,~F. Nanoscale Polymer Recognition by
  Spectral Signature in Scattering Infrared Near-Field Microscopy. \emph{Appl.
  Phys. Lett.} \textbf{2004}, \emph{85}, 5064--5066\relax
\mciteBstWouldAddEndPuncttrue
\mciteSetBstMidEndSepPunct{\mcitedefaultmidpunct}
{\mcitedefaultendpunct}{\mcitedefaultseppunct}\relax
\EndOfBibitem
\bibitem[Huth \latin{et~al.}(2012)Huth, Govyadinov, Amarie, Nuansing, Keilmann,
  and Hillenbrand]{Huth2012}
Huth,~F.; Govyadinov,~A.; Amarie,~S.; Nuansing,~W.; Keilmann,~F.;
  Hillenbrand,~R. Nano-FTIR Absorption Spectroscopy of Molecular Fingerprints
  at 20nm Spatial Resolution. \emph{Nano Lett.} \textbf{2012}, \emph{12},
  3973--3978\relax
\mciteBstWouldAddEndPuncttrue
\mciteSetBstMidEndSepPunct{\mcitedefaultmidpunct}
{\mcitedefaultendpunct}{\mcitedefaultseppunct}\relax
\EndOfBibitem
\bibitem[Liang \latin{et~al.}(1998)Liang, Hawthorne, and Swainson]{Liang1998}
Liang,~J.-J.; Hawthorne,~F.~C.; Swainson,~I.~P. Triclinic Muscovite: X-ray
  Diffraction, Neutron Diffraction and Photo-Acoustic FTIR Spectroscopy.
  \emph{Can. Mineral.} \textbf{1998}, \emph{36}, 1017--1027\relax
\mciteBstWouldAddEndPuncttrue
\mciteSetBstMidEndSepPunct{\mcitedefaultmidpunct}
{\mcitedefaultendpunct}{\mcitedefaultseppunct}\relax
\EndOfBibitem
\bibitem[Vedder(1964)]{Vedder1964}
Vedder,~W. Correlations Between Infrared Spectrum and Chemical Composition of
  Mica. \emph{Am. Mineral.} \textbf{1964}, \emph{49}, 736--768\relax
\mciteBstWouldAddEndPuncttrue
\mciteSetBstMidEndSepPunct{\mcitedefaultmidpunct}
{\mcitedefaultendpunct}{\mcitedefaultseppunct}\relax
\EndOfBibitem
\bibitem[Hale and Querry(1973)Hale, and Querry]{Hale1973}
Hale,~G.~M.; Querry,~M.~R. Optical Constants of Water in the 200-nm to
  200$\mu$-m Wavelength Region. \emph{Appl. Opt.} \textbf{1973}, \emph{12},
  555--563\relax
\mciteBstWouldAddEndPuncttrue
\mciteSetBstMidEndSepPunct{\mcitedefaultmidpunct}
{\mcitedefaultendpunct}{\mcitedefaultseppunct}\relax
\EndOfBibitem
\bibitem[Bertie \latin{et~al.}(1989)Bertie, Ahmed, and Eysel]{Bertie1989}
Bertie,~J.~E.; Ahmed,~M.~K.; Eysel,~H.~H. Infrared Intensities of Liquids. 5.
  Optical and Dielectric Constants, Integrated Intensities, and Dipole Moment
  Derivatives of Water and Water-d2 at 22.degree.C. \emph{J. Phys. Chem.}
  \textbf{1989}, \emph{93}, 2210--2218\relax
\mciteBstWouldAddEndPuncttrue
\mciteSetBstMidEndSepPunct{\mcitedefaultmidpunct}
{\mcitedefaultendpunct}{\mcitedefaultseppunct}\relax
\EndOfBibitem
\bibitem[Zhang \latin{et~al.}(2012)Zhang, Andreev, Fei, McLeod, Dominguez,
  Thiemens, Castro-Neto, Basov, and Fogler]{Zhang2012}
Zhang,~L.~M.; Andreev,~G.~O.; Fei,~Z.; McLeod,~A.~S.; Dominguez,~G.;
  Thiemens,~M.; Castro-Neto,~A.~H.; Basov,~D.~N.; Fogler,~M.~M. Near-Field
  Spectroscopy of Silicon Dioxide Thin Films. \emph{Phys. Rev. B}
  \textbf{2012}, \emph{85}, 075419\relax
\mciteBstWouldAddEndPuncttrue
\mciteSetBstMidEndSepPunct{\mcitedefaultmidpunct}
{\mcitedefaultendpunct}{\mcitedefaultseppunct}\relax
\EndOfBibitem
\bibitem[Govyadinov \latin{et~al.}(2014)Govyadinov, Mastel, Golmar, Chuvilin,
  Carney, and Hillenbrand]{Govyadinov2014}
Govyadinov,~A.~A.; Mastel,~S.; Golmar,~F.; Chuvilin,~A.; Carney,~P.~S.;
  Hillenbrand,~R. Recovery of Permittivity and Depth from Near-Field Data as a
  Step toward Infrared Nanotomography. \emph{ACS Nano} \textbf{2014}, \emph{8},
  6911--6921\relax
\mciteBstWouldAddEndPuncttrue
\mciteSetBstMidEndSepPunct{\mcitedefaultmidpunct}
{\mcitedefaultendpunct}{\mcitedefaultseppunct}\relax
\EndOfBibitem
\bibitem[Fei \latin{et~al.}(2011)Fei, Andreev, Bao, Zhang, S.~McLeod, Wang,
  Stewart, Zhao, Dominguez, Thiemens, Fogler, Tauber, Castro-Neto, Lau,
  Keilmann, and Basov]{Fei2011}
Fei,~Z.; Andreev,~G.~O.; Bao,~W.; Zhang,~L.~M.; S.~McLeod,~A.; Wang,~C.;
  Stewart,~M.~K.; Zhao,~Z.; Dominguez,~G.; Thiemens,~M. \latin{et~al.}
  Infrared Nanoscopy of Dirac Plasmons at the Graphene/SiO2 Interface.
  \emph{Nano Lett.} \textbf{2011}, \emph{11}, 4701--4705\relax
\mciteBstWouldAddEndPuncttrue
\mciteSetBstMidEndSepPunct{\mcitedefaultmidpunct}
{\mcitedefaultendpunct}{\mcitedefaultseppunct}\relax
\EndOfBibitem
\bibitem[McLeod \latin{et~al.}(2014)McLeod, Kelly, Goldflam, Gainsforth,
  Westphal, Dominguez, Thiemens, Fogler, and Basov]{McLeod2014}
McLeod,~A.~S.; Kelly,~P.; Goldflam,~M.~D.; Gainsforth,~Z.; Westphal,~A.~J.;
  Dominguez,~G.; Thiemens,~M.~H.; Fogler,~M.~M.; Basov,~D.~N. Model for
  Quantitative Tip-Enhanced Spectroscopy and the Extraction of
  Nanoscale-Resolved Optical Constants. \emph{Phys. Rev. B} \textbf{2014},
  \emph{90}, 085136\relax
\mciteBstWouldAddEndPuncttrue
\mciteSetBstMidEndSepPunct{\mcitedefaultmidpunct}
{\mcitedefaultendpunct}{\mcitedefaultseppunct}\relax
\EndOfBibitem
\bibitem[Gorshunov \latin{et~al.}(2013)Gorshunov, Zhukova, Torgashev, Lebedev,
  Shakurov, Kremer, Pestrjakov, Thomas, Fursenko, and Dressel]{Gorshunov2013}
Gorshunov,~B.~P.; Zhukova,~E.~S.; Torgashev,~V.~I.; Lebedev,~V.~V.;
  Shakurov,~G.~S.; Kremer,~R.~K.; Pestrjakov,~E.~V.; Thomas,~V.~G.;
  Fursenko,~D.~A.; Dressel,~M. Quantum Behavior of Water Molecules Confined to
  Nanocavities in Gemstones. \emph{J. Phys. Chem. Lett.} \textbf{2013},
  \emph{4}, 2015--2020\relax
\mciteBstWouldAddEndPuncttrue
\mciteSetBstMidEndSepPunct{\mcitedefaultmidpunct}
{\mcitedefaultendpunct}{\mcitedefaultseppunct}\relax
\EndOfBibitem
\bibitem[Mante \latin{et~al.}(2014)Mante, Chen, Wen, Chen, Yang, Huang,
  Ju~Chen, Chen, Gusev, Chen, Kuo, Sheu, and Sun]{Mante2014}
Mante,~P.-A.; Chen,~C.-C.; Wen,~Y.-C.; Chen,~H.-Y.; Yang,~S.-C.; Huang,~Y.-R.;
  Ju~Chen,~I.; Chen,~Y.-W.; Gusev,~V.; Chen,~M.-J. \latin{et~al.}  Probing
  Hydrophilic Interface of Solid/Liquid-Water by Nanoultrasonics. \emph{Sci.
  Rep.} \textbf{2014}, \emph{4}, 6249\relax
\mciteBstWouldAddEndPuncttrue
\mciteSetBstMidEndSepPunct{\mcitedefaultmidpunct}
{\mcitedefaultendpunct}{\mcitedefaultseppunct}\relax
\EndOfBibitem
\bibitem[Sechler \latin{et~al.}(2010)Sechler, DelSole, and De\'ak]{Sechler2010}
Sechler,~T.~D.; DelSole,~E.~M.; De\'ak,~J.~C. Measuring Properties of
  Interfacial and Bulk Water Regions in a Reverse Micelle with IR Spectroscopy:
  A Volumetric Analysis of the Inhomogeneously Broadened OH Band. \emph{J.
  Colloid Interface Sci.} \textbf{2010}, \emph{346}, 391--397\relax
\mciteBstWouldAddEndPuncttrue
\mciteSetBstMidEndSepPunct{\mcitedefaultmidpunct}
{\mcitedefaultendpunct}{\mcitedefaultseppunct}\relax
\EndOfBibitem
\bibitem[Algara-Siller \latin{et~al.}(2015)Algara-Siller, Lehtinen, Wang, Nair,
  Kaiser, Wu, Geim, and Grigorieva]{Algara-Siller2015}
Algara-Siller,~G.; Lehtinen,~O.; Wang,~F.~C.; Nair,~R.~R.; Kaiser,~U.;
  Wu,~H.~A.; Geim,~A.~K.; Grigorieva,~I.~V. Square Ice in Graphene
  Nanocapillaries. \emph{Nature} \textbf{2015}, \emph{519}, 443--445\relax
\mciteBstWouldAddEndPuncttrue
\mciteSetBstMidEndSepPunct{\mcitedefaultmidpunct}
{\mcitedefaultendpunct}{\mcitedefaultseppunct}\relax
\EndOfBibitem
\bibitem[Fei \latin{et~al.}(2012)Fei, Rodin, Andreev, Bao, McLeod, Wagner,
  Zhang, Zhao, Thiemens, Dominguez, Fogler, Neto, Lau, Keilmann, and
  Basov]{Fei2012}
Fei,~Z.; Rodin,~A.~S.; Andreev,~G.~O.; Bao,~W.; McLeod,~A.~S.; Wagner,~M.;
  Zhang,~L.~M.; Zhao,~Z.; Thiemens,~M.; Dominguez,~G. \latin{et~al.}
  Gate-Tuning of Graphene Plasmons Revealed by Infrared Nano-Imaging.
  \emph{Nature} \textbf{2012}, \emph{487}, 82--85\relax
\mciteBstWouldAddEndPuncttrue
\mciteSetBstMidEndSepPunct{\mcitedefaultmidpunct}
{\mcitedefaultendpunct}{\mcitedefaultseppunct}\relax
\EndOfBibitem
\bibitem[Mastel \latin{et~al.}(2015)Mastel, Govyadinov, de~Oliveira, Amenabar,
  and Hillenbrand]{Mastel2015}
Mastel,~S.; Govyadinov,~A.~A.; de~Oliveira,~T. V. A.~G.; Amenabar,~I.;
  Hillenbrand,~R. Nanoscale-Resolved Chemical Identification of Thin Organic
  Films Using Infrared Near-Field Spectroscopy and Standard Fourier Transform
  Infrared References. \emph{Appl. Phys. Lett.} \textbf{2015}, \emph{106},
  023113\relax
\mciteBstWouldAddEndPuncttrue
\mciteSetBstMidEndSepPunct{\mcitedefaultmidpunct}
{\mcitedefaultendpunct}{\mcitedefaultseppunct}\relax
\EndOfBibitem
\bibitem[Zhou \latin{et~al.}(2013)Zhou, Hu, Wang, Xu, Wang, Bai, Shan, and
  Lu]{Zhou2013}
Zhou,~Z.; Hu,~Y.; Wang,~H.; Xu,~Z.; Wang,~W.; Bai,~X.; Shan,~X.; Lu,~X. DNA
  Translocation through Hydrophilic Nanopore in Hexagonal Boron Nitride.
  \emph{Sci. Rep.} \textbf{2013}, \emph{3}, 3287\relax
\mciteBstWouldAddEndPuncttrue
\mciteSetBstMidEndSepPunct{\mcitedefaultmidpunct}
{\mcitedefaultendpunct}{\mcitedefaultseppunct}\relax
\EndOfBibitem
\bibitem[Jariwala \latin{et~al.}(2014)Jariwala, Sangwan, Lauhon, Marks, and
  Hersam]{Jariwala2014}
Jariwala,~D.; Sangwan,~V.~K.; Lauhon,~L.~J.; Marks,~T.~J.; Hersam,~M.~C.
  Emerging Device Applications for Semiconducting Two-Dimensional Transition
  Metal Dichalcogenides. \emph{ACS Nano} \textbf{2014}, \emph{8},
  1102--1120\relax
\mciteBstWouldAddEndPuncttrue
\mciteSetBstMidEndSepPunct{\mcitedefaultmidpunct}
{\mcitedefaultendpunct}{\mcitedefaultseppunct}\relax
\EndOfBibitem
\bibitem[Li \latin{et~al.}(2014)Li, Yu, Ye, Ge, Ou, Wu, Feng, Chen, and
  Zhang]{Li2014}
Li,~L.; Yu,~Y.; Ye,~G.~J.; Ge,~Q.; Ou,~X.; Wu,~H.; Feng,~D.; Chen,~X.~H.;
  Zhang,~Y. Black Phosphorus Field-Effect Transistors. \emph{Nat. Nanotechnol.}
  \textbf{2014}, \emph{9}, 372--377\relax
\mciteBstWouldAddEndPuncttrue
\mciteSetBstMidEndSepPunct{\mcitedefaultmidpunct}
{\mcitedefaultendpunct}{\mcitedefaultseppunct}\relax
\EndOfBibitem
\bibitem[Wood \latin{et~al.}(2014)Wood, Wells, Jariwala, Chen, Cho, Sangwan,
  Liu, Lauhon, Marks, and Hersam]{Wood2014}
Wood,~J.~D.; Wells,~S.~A.; Jariwala,~D.; Chen,~K.-S.; Cho,~E.; Sangwan,~V.~K.;
  Liu,~X.; Lauhon,~L.~J.; Marks,~T.~J.; Hersam,~M.~C. Effective Passivation of
  Exfoliated Black Phosphorus Transistors against Ambient Degradation.
  \emph{Nano Lett.} \textbf{2014}, \emph{14}, 6964--6970\relax
\mciteBstWouldAddEndPuncttrue
\mciteSetBstMidEndSepPunct{\mcitedefaultmidpunct}
{\mcitedefaultendpunct}{\mcitedefaultseppunct}\relax
\EndOfBibitem
\bibitem[Dai \latin{et~al.}(2015)Dai, Ma, Andersen, Mcleod, Fei, Liu, Wagner,
  Watanabe, Taniguchi, Thiemens, Keilmann, Jarillo-Herrero, Fogler, and
  Basov]{Dai2015}
Dai,~S.; Ma,~Q.; Andersen,~T.; Mcleod,~A.~S.; Fei,~Z.; Liu,~M.~K.; Wagner,~M.;
  Watanabe,~K.; Taniguchi,~T.; Thiemens,~M. \latin{et~al.}  Subdiffractional
  Focusing and Guiding of Polaritonic Rays in a Natural Hyperbolic Material.
  \emph{Nat. Commun.} \textbf{2015}, \emph{6}, --\relax
\mciteBstWouldAddEndPuncttrue
\mciteSetBstMidEndSepPunct{\mcitedefaultmidpunct}
{\mcitedefaultendpunct}{\mcitedefaultseppunct}\relax
\EndOfBibitem
\bibitem[Ocelic \latin{et~al.}(2006)Ocelic, Huber, and Hillenbrand]{Ocelic2006}
Ocelic,~N.; Huber,~A.; Hillenbrand,~R. Pseudoheterodyne Detection for
  Background-free Near-Field Spectroscopy. \emph{Appl. Phys. Lett.}
  \textbf{2006}, \emph{89}, 101124\relax
\mciteBstWouldAddEndPuncttrue
\mciteSetBstMidEndSepPunct{\mcitedefaultmidpunct}
{\mcitedefaultendpunct}{\mcitedefaultseppunct}\relax
\EndOfBibitem
\bibitem[Gambetta \latin{et~al.}(2008)Gambetta, Ramponi, and
  Marangoni]{Gambetta2008}
Gambetta,~A.; Ramponi,~R.; Marangoni,~M. Mid-Infrared Optical Combs from a
  Compact Amplified Er-Doped Fiber Oscillator. \emph{Opt. Lett.} \textbf{2008},
  \emph{33}, 2671--2673\relax
\mciteBstWouldAddEndPuncttrue
\mciteSetBstMidEndSepPunct{\mcitedefaultmidpunct}
{\mcitedefaultendpunct}{\mcitedefaultseppunct}\relax
\EndOfBibitem
\bibitem[Herzinger \latin{et~al.}(1998)Herzinger, Johs, McGahan, Woollam, and
  Paulson]{Herzinger1998}
Herzinger,~C.~M.; Johs,~B.; McGahan,~W.~A.; Woollam,~J.~A.; Paulson,~W.
  Ellipsometric Determination of Optical Constants for Silicon and Thermally
  Grown Silicon Dioxide \textit{via} a Multi-Sample, Multi-Wavelength,
  Multi-Angle Investigation. \emph{J. Appl. Phys.} \textbf{1998}, \emph{83},
  3323--3336\relax
\mciteBstWouldAddEndPuncttrue
\mciteSetBstMidEndSepPunct{\mcitedefaultmidpunct}
{\mcitedefaultendpunct}{\mcitedefaultseppunct}\relax
\EndOfBibitem
\end{mcitethebibliography}

\includepdf[pages={-}]{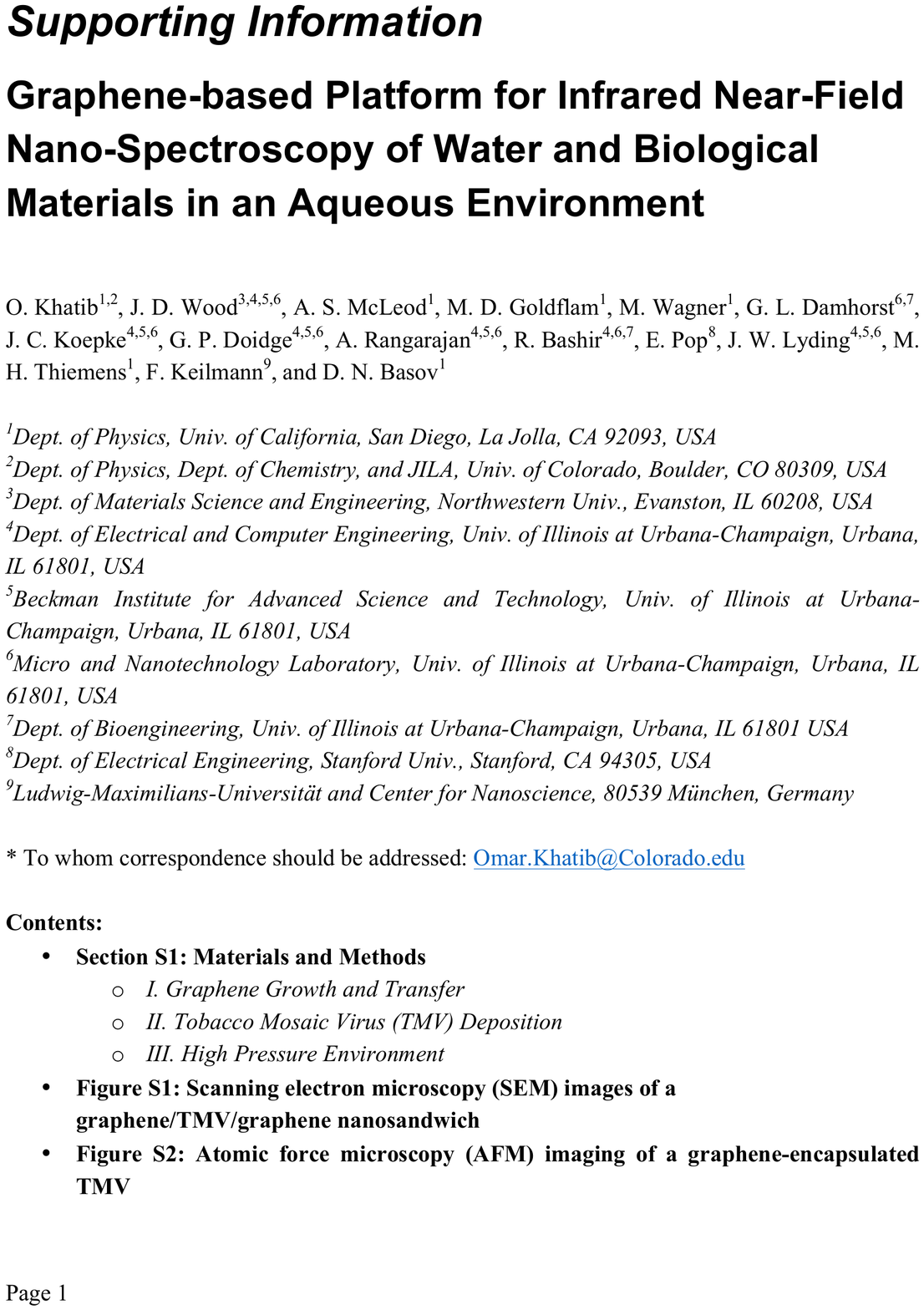}

\end{document}